\documentclass[mathleft,fleqn,%
]{an}
\usepackage{amsmath}
\usepackage{graphicx}
\usepackage[varg]{txfonts}
\usepackage{natbib}
\bibpunct{(}{)}{;}{a}{}{,}
\begin{document}

\Pagespan{1}{}
\Yearpublication{2016}%
\Yearsubmission{2016}%
\Month{0}%
\Volume{999}%
\Issue{0}%
\DOI{asna.201400000}%

\title{Magnetic field reconstruction based on sunspot oscillations}

\author{J. L\"{o}hner-B\"{o}ttcher \fnmsep\thanks{Corresponding author:
        {jlb@kis.uni-freiburg.de}}
\and  N. Bello Gonz\'{a}lez
\and  W. Schmidt
}
\titlerunning{Magnetic field reconstruction based on sunspot oscillations}
\authorrunning{J. L\"{o}hner-B\"{o}ttcher et al.}
\institute{
Kiepenheuer-Institut f\"{u}r Sonnenphysik, Sch\"{o}neckstrasse 6, 79104 Freiburg, Germany}

\received{XXXX}
\accepted{XXXX}
\publonline{XXXX}


\abstract{The magnetic field of a sunspot guides magnetohydrodynamic waves toward higher atmospheric layers. In the upper photosphere and lower chromosphere, wave modes with periods longer than the acoustic cut-off period become evanescent. The cut-off period essentially changes due to the atmospheric properties, e.g., increases for larger zenith inclinations of the magnetic field. In this work, we aim at introducing a novel technique of reconstructing the magnetic field inclination on the basis of the dominating wave periods in the sunspot chromosphere and upper photosphere. On 2013 August 21st, we observed an isolated, circular sunspot (NOAA11823) for 58\,min in a purely spectroscopic multi-wavelength mode with the Interferometric Bidimensional Spectro-polarimeter (IBIS) at the Dunn Solar Telescope. By means of a wavelet power analysis, we retrieved the dominating wave periods and reconstructed the zenith inclinations in the chromosphere and upper photosphere. The results are in good agreement with the lower photospheric HMI magnetograms. The sunspot's magnetic field in the chromosphere inclines from almost vertical (0$^{\circ}$) in the umbra to around 60$^{\circ}$ in the outer penumbra. With increasing altitude in the sunspot atmosphere, the magnetic field of the penumbra becomes less inclined. We conclude that the reconstruction of the magnetic field topology on the basis of sunspot oscillations yields consistent and conclusive results. The technique opens up a new possibility to infer the magnetic field inclination in the solar chromosphere.}

\maketitle

\section{Introduction}\label{sec_intro}
Sunspot waves are one of the most prominent and dynamical phenomena in the solar atmosphere. In the sunspot chromosphere, umbral flashes \citep{1969SoPh....7..351B} and running penumbral waves \citep{1972SoPh...27...71G,1972ApJ...178L..85Z} appear as unceasing brightness and velocity oscillations. Numerous observational studies \citep[e.g.,][]{2006ApJ...640.1153C,2007ApJ...671.1005B,2015ApJ...800..129M} confirmed that sunspot waves are propagating slow-mode magnetoacoustic waves which are guided upward along the magnetic field lines. From the umbra to the outer penumbra, the wave characteristics change owing to the increase in inclination of the magnetic waveguides against the zenith. 

As the travel distance to a distinct chromospheric sampling layer increases with the magnetic field inclination, the penumbral waves delay and seem to follow the umbral waves. The second major difference between both phenomena is the wave periodicity (see Fig.\,\ref{fig_sunspot_context}\,d). Whereas the umbral chromosphere is dominated by 2.5--3\,min waves, the peak periods of penumbral waves increase as a function of radial distance from 3\,min at the umbral-penumbral boundary up to 8\,min at the outer penumbra \citep{2013ApJ...779..168J}. 

This change in wave propagation results from the acoustic cut-off layer. Wave modes with periods longer than the characteristic cut-off period become evanescent. For vertical magnetic fields, the cut-off period scales proportionally with the local speed of sound or atmospheric temperature. The typical value for sunspot umbrae ranges around 192\,s. As proposed by \citet{1977A&A....55..239B} and confirmed by, e.g., \citet{2004Natur.430..536D}, the inclination of the magnetic field lines against the zenith strongly impacts the acoustic cut-off period. The acoustic cut-off period
\begin{equation}
\hspace{2.4cm} T_{\rm  cut,\,\Phi_{\rm B}} = \frac{4\pi\  c_{\rm S}}{\gamma\ g\ \cos\Phi_{\rm B}}\,,
\label{eq_peak_period_new}
\end{equation}
effectively depends on the sound speed $c_{\rm S}$ and the zenith inclination $\Phi_{\rm B}$ of the magnetic field. The constant adiabatic index $\gamma = 5/3$ (monoatomic gases) and the solar gravitational acceleration $g\approx274\,\rm{m/s}$ enter the equation. In the power spectrum, those wave modes with periods slightly shorter than the critical cut-off value dominate. The empirical formula $T_{\rm  cut,\Phi_{\rm B}} = m\ T_{\rm PEAK}$ with the factor $m=1.25$ often describes a good indicator of the the cut-off period from the measured peak period $T_{\rm PEAK}$ in the power spectrum \citep{2006A&A...456..689T,2006RSPTA.364..313B}. Transposing and inverting Eq.\,(\ref{eq_peak_period_new}), and substituting the adiabatic sound speed yields the zenith inclination of the magnetic field as
\begin{equation}
\hspace{1.9cm}\Phi_{\rm B} = \cos^{-1}\left(\frac{4\pi\ \sqrt{\frac{\gamma\ R\ \vartheta_{\rm K}}{M}}}{m\ \gamma\  g\ T_{\rm PEAK}}\right)\,,
\label{eq_field_inclination}
\end{equation}
with the peak periods $T_{\rm PEAK}$ in seconds, the atmospheric temperature $\vartheta_{\rm K}$ in Kelvin, and the specific gas constant $R/M$ with the mean molar mass $M$ and gas constant $R$.

In the following, the novel method of reconstructing the magnetic field inclination from the atmospheric and oscillatory characteristics is applied to the observations which are presented in Section \ref{sec_data}. The determination of the two crucial components, the dominating wave period $T_{\rm PEAK}$ and the atmospheric temperature $\vartheta_{\rm K}$, is described. In Section \ref{sec_results}, the most fundamental results for upper photospheric and lower chromospheric zenith inclination are discussed. The conclusions are drawn in Section \ref{sec_conclusions}.

\section{Observations and analysis}\label{sec_data}
\begin{figure*}[htbp]
\begin{center}
\includegraphics[trim=0.2cm 0cm 0cm 0cm,clip,width=16.5cm]{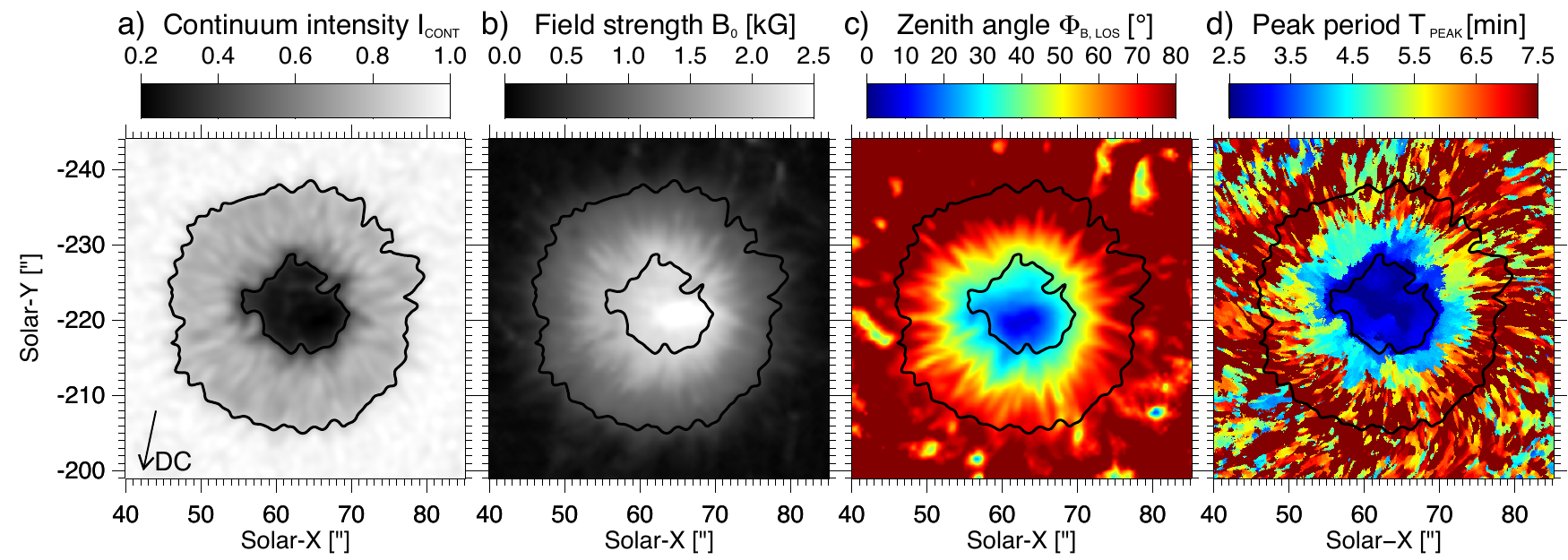}
\caption{Time-averaged (1h) physical quantities of sunspot NOAA11823. The sunspot is displayed in 
\textbf{a)} normalized HMI continuum intensities $I_{\rm CONT}$ at around 617\,nm, 
\textbf{b)} absolute magnetic field strengths $B_0$ and 
\textbf{c)} line-of-sight magnetic field inclinations $\Phi_{\rm B, LOS}$, both from HMI inversions of the polarimetric signals of \ion{Fe}{I} $617.33\,\rm{nm}$. The photospheric inclination angles are scaled from the line-of-sight direction of $0^{\circ}$ (blue) to a roughly perpendicular orientation of $80^\circ$ to the line-of-sight (red). \textbf{d)}: The chromospheric distribution of dominant sunspot waves is shown by the peak periods $T_{\rm PEAK}$ from IBIS intensity oscillations of the \ion{Ca}{II} $854.2\,\rm{nm}$ line core. The periods are scaled from 2.5\,min (blue) to 7.5\,min (red). The black contours mark the umbral (inner) and penumbral (outer) boundaries from continuum intensity. The arrow (in panel a) is pointing toward disk center (DC).}
\label{fig_sunspot_context}
\includegraphics[trim=0cm 0cm 0.0cm -1cm,clip,width=16.5cm]{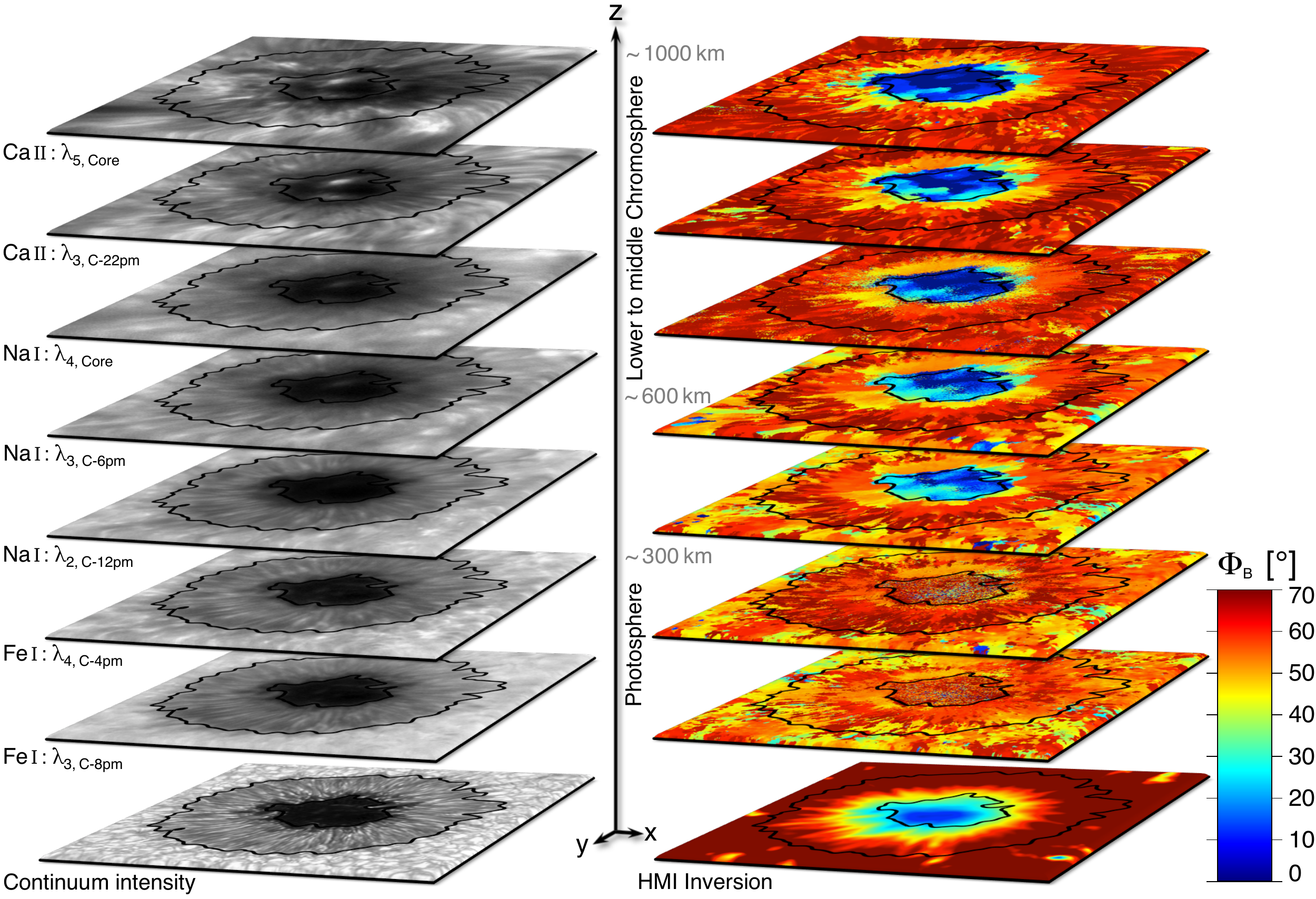}
\caption{Sunspot NOAA11823 in a three-dimensional view of intensities and reconstructed zenith inclinations $\Phi_{\rm B}$ of the magnetic field from photospheric to chromospheric heights. The intensities on the left show the sunspot at 2013 August 21 at 15:00:06UTC at several line core and wing positions of \ion{Fe}{I} $630.15\,\rm{nm}$, \ion{Na}{I} $589.59\,\rm{nm,}$ and \ion{Ca}{II} $854.21\,\rm{nm}$. A bright umbral flash is present in the central umbra. The inferred zenith inclinations on the right are scaled from $0^\circ$ (dark blue) to $70^\circ$ (dark red). The sunspot boundaries from continuum intensity (bottom left) are contoured in black. The photospheric line-of-sight inclination from the HMI inversion is added for comparison (bottom right). The image positions along the z-axis are ordered at a non-equidistant scale according to their estimated height above the optical depth unity.}
\label{fig_field_inclination_3D}
\end{center}
\end{figure*}

The isolated fully-developed sunspot of active region NOAA11823 was observed on August 21st 2013 from 14:53UTC to 15:51UTC with the Dunn Solar Telescope (DST) at the National Solar Observatory in New Mexico. In addition to the spectroscopic observations which were performed with the Interferometric BIdimensional Spectro-polarimeter \citep[IBIS;][]{2006SoPh..236..415C}, spectro-polarimetric data were acquired from the Helioseismic Magnetic Imager (HMI) aboard the Solar Dynamics Observatory. The sunspot in time-averaged continuum intensity is shown in Fig.\,\ref{fig_sunspot_context} (panel a) and Fig.\,\ref{fig_field_inclination_3D} (bottom left). The spot was located close to the solar disk center at $(X,Y)=(63\arcsec,-222\arcsec)$ with a heliocentric angle of $\theta = 14^\circ$, has a circular penumbra and a diameter of 24\,Mm. 

To gain knowledge and context information about the photospheric magnetic field vector, the Very Fast Inversion of the Stokes Vector \citep[VFISV;][]{2011SoPh..273..267B} was applied on the polarimetric signals of the \ion{Fe}{I} $617.33\,\rm{nm}$ line of HMI. As shown in Fig.\,\ref{fig_sunspot_context} (panels b and c), the umbral core yields an absolute magnetic field strength of up to 2.6\,kG and an inclination $\Phi_{\rm B,\,LOS}$ against the line-of-sight of less than 20$^{\circ}$. According to the apparent displacement of the line-of-sight magnetic field toward the disc center, orientation in the umbra can be assumed as vertical. Toward the outer penumbra, the magnetic field strength decreases almost linearly (see Fig.\,\ref{fig_sunspot_parameters}\,b). 

The spectrometric multi-wavelength IBIS observations sampled various non-equidistant line core and wing positions of the \ion{Na}{I}\,D1 $589.59\,\rm{nm}$ and \ion{Ca}{II} $854.21\,\rm{nm}$ lines, covering the photosphere and chromosphere up to an atmospheric formation height of around 1000\,km above the wavelength dependent optical depth unity. In addition, the \ion{Fe}{I} $630.15\,\rm{nm}$ line was observed to gain information about the lower photosphere. With an overall cadence of $13.2\,\rm{s}$, the Nyquist criterion enables the investigation of oscillations with periods as small as $26.5\,\rm{s}$. To obtain the dominating wave periods across the sunspot atmosphere, the spectral intensities at the wavelength positions were analyzed with wavelet techniques \citep{1998BAMS...79...61T}. Finally, the dominating wave period  $T_{\rm PEAK}$ was extracted from the global wavelet power spectra as the position of the maximum below 10\,min. Exemplarily, the distribution of peak periods in the sunspot chromosphere (line core of \ion{Ca}{II} $854.21\,\rm{nm}$) is shown in Fig.\,\ref{fig_sunspot_context} (panel d). Whereas the umbral area is characterized by 2.5--3\,min periods, the penumbra yields a filamentary radial increase of wave periods up to 8\,min at the outer penumbra and beyond. The three-dimensional evolution of wave periods is illustrated in Fig.\,2 of \citet{2015A&A...580A..53L}.

Besides the peak period $T_{\rm PEAK}$, which is the most sensitive parameter for the computation of the field inclination in Eq.\,(\ref{eq_field_inclination}), the local sound speed $c_{\rm S}$ is the second crucial component to be evaluated. A simple approach was chosen to estimate the local temperature $\vartheta_{\rm K}$ in the sunspot atmosphere. First, a reference position within the sunspot region is defined, either in the central core of the umbra or in the quiet sun surrounding the sunspot.  According to the position, the corresponding semi-empirical atmospheric temperature model from \citet{1986ApJ...306..284M} is selected. Then, the formation height of the spectral position is estimated by means of spectral contribution functions \citep{2008A&A...480..515C,2010ApJ...709.1362L}. The model temperature is extracted and averaged for a layer centered at the respective estimated formation height ($\pm100\,{\rm km\,s^{-1}}$). Finally, according to Plancks law and the Wien approximation, the atmospheric temperature is calculated on the basis of the relative spectral brightness across the sunspot region. Based on an umbral and quiet sun model, both results are in good agreement and enter the reconstruction equation of the magnetic field inclination. This simple approach return a suited estimation of the atmospheric temperature and local sound speed. To evaluate the atmospheric properties in an even preciser way, a pixel-wise inversion of the spectral line profile can and , in the future, will be applied to the methodology.

\section{Results and discussion}\label{sec_results}
The zenith inclination of the sunspot's magnetic field is reconstructed successfully by Eq.\,(\ref{eq_field_inclination}) on the basis of the atmospheric and oscillatory parameters. According to the photospheric HMI magnetic field inclination and an additional coronal magnetic field extrapolation with a Potential-Field Source-Surface \citep[PFSS;][]{2003SoPh..212..165S} model, a vertical magnetic field can be assumed for the central umbra. To maintain the vertical orientation of the central umbra throughout the atmosphere, the peak-to-cut-off factor $m$ of Eq.\,(\ref{eq_field_inclination}) has to be adapted to the atmospheric layer, from 1.1 in the middle photosphere to 1.25 in the middle chromosphere. 

\begin{figure}[htbp]
\begin{center}
\includegraphics[trim=0.0cm 0cm 8.8cm 0cm,clip,height=5.0cm]{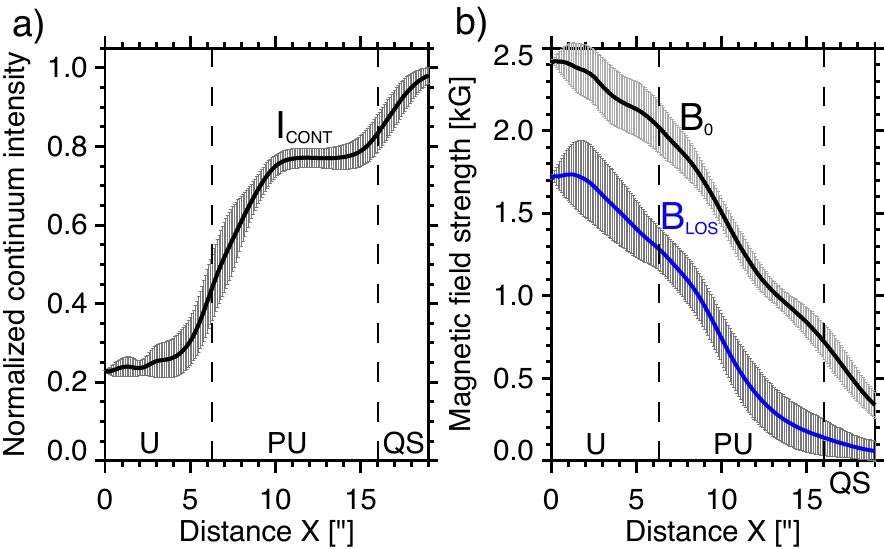}\\
\hspace{0.1cm}\includegraphics[trim=9.0cm 0cm 0cm 0cm,clip,height=5.1cm]{7554_reconstruction_field_parameter_comparison.pdf}
\caption{Azimuthally averaged sunspot parameters as a function of radial distance $X$ from the sunspot barycenter: \textbf{a)} the normalized continuum intensities at around 617\,nm, \textbf{b)} absolute ($B_0$, black curve) and line-of-sight ($B_{\rm LOS}$, blue curve) magnetic field strengths from photospheric HMI inversions, \textbf{c)} zenith inclinations $\Phi_{\rm B}$ of the magnetic field in the lower photosphere (black curve) and chromosphere (blue curve), and \textbf{d)} peak wave periods $T_{\rm PEAK}$ in the middle photosphere (black curve) and chromosphere (blue curve). The average sunspot boundaries are marked by the vertical dashed lines. The standard deviations along the azimuth are plotted as error bars. 
}
\label{fig_sunspot_parameters}
\end{center}
\end{figure}

\begin{figure}[htbp]
\begin{center}
\includegraphics[trim=0cm 0cm 0cm 0cm,clip,width=8.3cm]{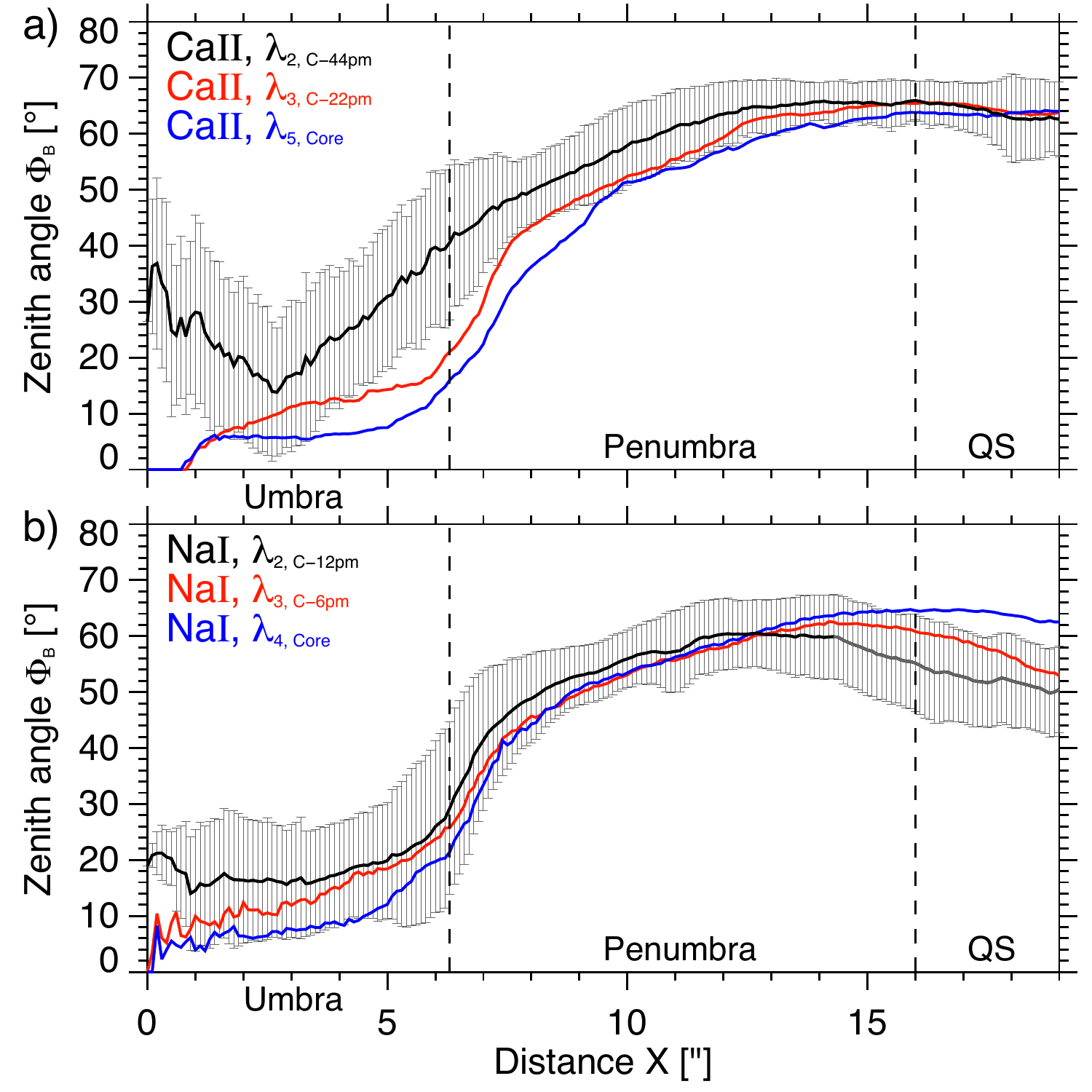}
\caption{Zenith inclinations $\Phi_{\rm B}$ of the sunspot's magnetic field from the photosphere to the chromosphere. The azimuthally averaged inclination angles are plotted as a function of radial distance $X$ from the sunspot barycenter. The panels show the evolution of the inclination with increasing atmospheric height from the middle photosphere (black curves) to the lower to middle chromosphere (red and blue curves). The curves stem from distinct wavelength positions: \textbf{a)} \ion{Ca}{II} 854.21\,nm at the blue line wing (black curve, $\approx250$\,km), blue line core (red curve, $\approx900$\,km), and line minimum (blue curve, $\approx1000$\,km); \textbf{b)} \ion{Na}{I} 589.59\,nm at the blue line wing (black curve, $\approx300$\,km), blue line core (red curve, $\approx600$\,km), and line minimum (blue curve, $\approx1000$\,km).  The error bars to the black curve are the standard deviations along the azimuth. The vertical dashed lines mark the average sunspot boundaries.}
\label{fig_sunspot_azimuthal_average}
\end{center}
\end{figure}

The resulting three-dimensional distribution of the sunspot's magnetic field inclination is illustrated in Fig.\,\ref{fig_field_inclination_3D} on the right. The spectral intensities at the corresponding wavelength positions of the \ion{Fe}{I}, \ion{Na}{I} D1, and \ion{Ca}{II} line are ordered according to their estimated formation height and plotted on the left. Except for the lower photosphere (two \ion{Fe}{I} line core positions) in which the acoustic cut-off not yet defines the dominating wave propagation, the results especially in the chromosphere and upper photosphere are in good agreement with the lower photospheric HMI line-of-sight magnetic field inclination on the bottom right. The zenith inclinations of the sunspot's magnetic field increases systematically from a vertical orientation (0$^{\circ}$) in the umbra to around 60$^{\circ}$ in the outer penumbra. The penumbral magnetic field inclination decreases with increasing altitude. It is best reflected in Fig.\,\ref{fig_sunspot_parameters}\,c  and Fig.\,\ref{fig_sunspot_azimuthal_average} in which the azimuthally averaged (including spherical projection effects) magnetic field inclinations in the photosphere (black curves) and chromosphere (blue and red curves) are plotted as a function of radial distance from the sunspot barycenter. The robustness of the methodology is verified by the consistency of the results from both chromospheric spectral lines. The decrease in magnetic field inclination with height is in line with the topology of a magnetic field configuration that fans out toward the less dense upper atmosphere \citep{2001ApJ...547.1130W}. 

\section{Conclusions}\label{sec_conclusions}
The reconstruction of the magnetic field inclination on the basis of sunspot oscillations from spectroscopic observations yields inherently consistent and conclusive results. Since the effect of the acoustic cut-off layer on wave propagation is employed, the reconstruction yields the best results in the chromosphere and upper photosphere. This novel technique opens up a new possibility to infer the magnetic field inclination in the solar chromosphere and provides a promising alternative to the investigation of spectro-polarimetric full-Stokes signals in the higher atmosphere. 
%



\acknowledgements
The data were acquired in service mode operation within the transnational ACCESS program of SOLARNET, an EU-FP7 integrated activity project. The IBIS instrument at the DST (NSO) was operated by INAF personnel, with special thanks to Gianna Cauzzi. HMI data were used by courtesy of NASA/SDO and HMI science teams. This work was prepared within the Centre for Advanced Solar Spectro-polarimetric Data Analysis (CASSDA) project, funded by the Senatsausschuss of the Leibniz Association, Ref.-No. SAW-2012-KIS-5. 
 
\bibliographystyle{an}
\bibliography{articles}


\end{document}